\documentclass{article}
\usepackage[dvips]{graphicx}
\usepackage{subfigure}
\usepackage{wrapfig}
 \usepackage{setspace}
\begin{document}
\oddsidemargin -0.31in
\title{The sensitivity of the population of states to the value of $q$ and the legitimate range of $q$ in Tsallis statistics}

\author{Ali M. Nassimi$^{a,b}$ and  Gholamabbas Parsafar$^b$ \\
$^a$Department of Chemistry, University of
Toronto, Toronto, ON, M5S 3H6, Canada.\\$^b$Department of Chemistry, Sharif University of Technology, Tehran, 11365-9516 Iran.}
\maketitle
\bibliographystyle{unsrt}

\begin{abstract}
In the framework of the Tsallis statistical mechanics, for the
spin-$\frac{1}{2}$ and the harmonic oscillator, we study the
change of the population of states when the parameter $q$ is
varied; the results show that the difference between predictions
of the Boltzmann--Gibbs and Tsallis Statistics can be much smaller
than the precision of any existing experiment. Also, the relation
between the privilege of rare/frequent event and the value of $q$
is restudied. This relation is shown to be more complicated than
the common belief about it. Finally, the convergence criteria of
the partition function of some simple model systems, in the
framework of Tsallis Statistical Mechanics, is studied; based on
this study , we conjecture that $q \leq 1$, in the thermodynamic
limit.
\subsection*{Keywords}
\it{Non-extensive statistical mechanics}, \it{Tsallis statistics},
\it{legitimate range of $q$}, \it{privilege of rare events}, and
\it{sensitivity to the value of $q$}.
\end{abstract}
%\footenotetext{email: \it{ali.nassimi@utoronto.ca}}
\section{Introduction}
The Boltzmann--Gibbs (BG) entropy is defined as
\begin{equation}
S=-k_B \sum_i p_i \ln p_i \label{Boltzmann-entropy},
\end{equation}
where $p_i$ is the probability of finding the system in the state
i and $k_B$ is the Boltzmann constant. According to the
information theory-based formulation of Statistical Mechanics, we
can consider the appropriate constraints for each ensemble and
derive the probability of having the system in each of its states
by finding the extremum of the entropy, (\ref{Boltzmann-entropy}),
\cite{katz}.

A generalized form for the entropy is \cite{tsallis}
\begin{equation}\label{Tsallis-entropy}
S_q=k \frac{1-\sum_{i=1}^W p_i^q}{q-1},
\end{equation}
where $q$ is the nonextensivity index, and k is a
constant. Statistical Mechanics is generalized, by finding the
extremum of (\ref{Tsallis-entropy}) instead of
(\ref{Boltzmann-entropy}). The result is called \it{Nonextensive Statistical Mechanics} or \it{Tsallis Statistics}.
Equation (\ref{Tsallis-entropy}) goes to equation (\ref{Boltzmann-entropy})
in the limit of $q \rightarrow 1$; also, every relation in this
new statistics goes to its corresponding relation in the BG
statistics, in the limit of $q \rightarrow 1$ \cite{abe:okamoto}.
The distribution functions arising in these statistics have found
wide applications through sciences which were commonly considered
to be out of the realm of Statistical Mechanics
\cite{gellman:tsallis}. The $q$-expectation value of an operator
$A$ is defined through $\langle A \rangle_q=\sum_{i=1}^W p_i^q
A_i$, where $A_i$ represent the value of the observable $A$, when the
system is in the state i; this definition is replaced for the usual
expectation value relation $\langle A \rangle =\sum_ip_iA_i$ in
the BG statistics. It is claimed that, systems containing
long-range interactions and/or long-range
microscopic memory (i.e., non-Markovian processes) have to be
described by Tsallis Statistics.

The normalization condition and the energy constraint of the canonical ensemble in the BG statistics are, respectively,
\begin{equation}\label{BGconstraint}
\sum_{i=1}^W p_i = 1 \hspace{1.5cm} \rm{and} \hspace{1.5cm} \sum_{i=1}^W p_i \epsilon_i = U,
\end{equation}
where $\epsilon_i$ represents the energy of the system
in its i'th microstate. While, the normalization condition is generaly accepted, the energy constraint is somehow ambiguous in this
generalization. First, it has been considered to be the same as
(\ref{BGconstraint}) \cite{tsallis}, this assumption yields
\begin{equation}\label{first tsallis probability}
p_i=\big[(1-q)(\alpha+\beta \epsilon_i)/q \big]^{1/(1-q)},
\end{equation}
where $\alpha$ and $\beta$  are undetermined Lagrange multipliers.
The position of the Lagrange multiplier $\alpha$  makes it
difficult to find its value by using the normalization condition in
(\ref{BGconstraint}). Thus, Curado and Tsallis suggest \cite{curado:tsallis}
\begin{equation}\label{CTconstraint}
\sum_{i=1}^Wp_i^q \epsilon_i=U_q
\end{equation}
as the energy constraint, which results respectively in the following probability and partition function,
\begin{equation}\label{tsallis probability}
p_i=(Z_q)^{-1}\big[1-(1-q) \beta \epsilon_i \big]^{1/(1-q)}
\hspace{1cm} \rm{and} \hspace{1 cm}
Z_q=\sum_{i=1}^W \big[1-(1-q) \beta \epsilon_i \big]^{1/(1-q)}.
\end{equation}
There are more complex proposals (e.g. \cite{tsallis:mendes});
but, it is shown that these versions of $p_i$ and $Z_q$ are all
equivalent to each other. They can be transformed to each other by
the appropriate change of variable, i.e., $\beta \rightarrow
\beta'$ \cite{Ferri:martinez}. It should be mentioned that wherever the expression in square brackets is
negative $p_i = 0$ by postulate.

We can ask whether it is
possible for a system yielding the same data either with a value
of $q$ not equal to one or with the BG statistics. Thus, in section
(\ref{Sensitivity}), we study the sensitivity of the population of
states to the value of $q$. The effect of the parameter $q$ on the weight of rare and
frequent events will be addressed in section (\ref{Rare}). The beauty of the Statistical-Mechanics is in
evaluating macroscopic properties from microscopic properties. But in
the non-extensive formalism, we need to know the value of $q$ in
addition to the microscopic properties. Although, there is no general way for evaluating $q$ a priori.
But, the possibility of confining the range of possible values of $q$ will be addressed in section \ref{legitimate}.

\section{Sensitivity of the population of states to the value of $q$} \label{Sensitivity}
\begin{wrapfigure}{r}{80mm}
  \begin{center}
     \includegraphics*[viewport=72 503 335 705, scale=0.90]{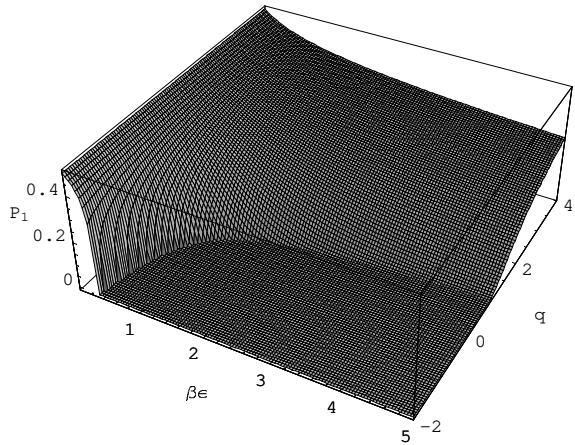}
   \end{center}
  \caption{The probability of a two-state system being in the higher energy state, versus $q$ and $\beta \epsilon$.}
  \label{fig1}
\end{wrapfigure}

The population of states in a two-state system (TSS) with energies 0 and
$\epsilon$ are, respectively,
\begin{equation}\label{population0}
P_0=\frac{1}{1+[1-(1-q)\beta \epsilon]^{\frac{1}{1-q}}},
\hspace{0.75cm} \rm{and} \hspace{0.75cm} P_1 = \frac{[1- (1- q)\beta \epsilon]^\frac{1}{1- q}}{1+[1-(1- q) \beta \epsilon]^\frac{1}{1-q}}.
\end{equation}
Substituting equation (\ref{population0}) into equation (\ref{Tsallis-entropy}) yields
\begin{equation}
\label{entropy2}
S_q=\frac{-1+\{1+[1-(1-q)\beta \epsilon]^\frac{1}{1-q}\}^{-q}+\{1+[1-(1-q)\beta \epsilon]^\frac{1}{q-1}\}^{-q}}{1-q}.
\end{equation}
Because of the form of these equations, it is difficult to study
their behavior analytically. $P_1$ versus $q$ and $\beta \epsilon
$ have been sketched in figure (\ref{fig1}). At constant values of
$q$, we can see the increase of $P_1$ toward 0.5 by
decreasing the value of $\beta \epsilon$, as expected. At constant
$\beta \epsilon$, it is seen that $q$ is playing a role similar to
 the temperature. For a sane study we should first estimate physical value of $\beta \epsilon
$.

\begin{figure}
 \begin{center}
  \subfigure[at the value of $\beta \epsilon=0.01$]{\label{excitedprobability-a} \includegraphics*[viewport=80 565 310  708,clip]{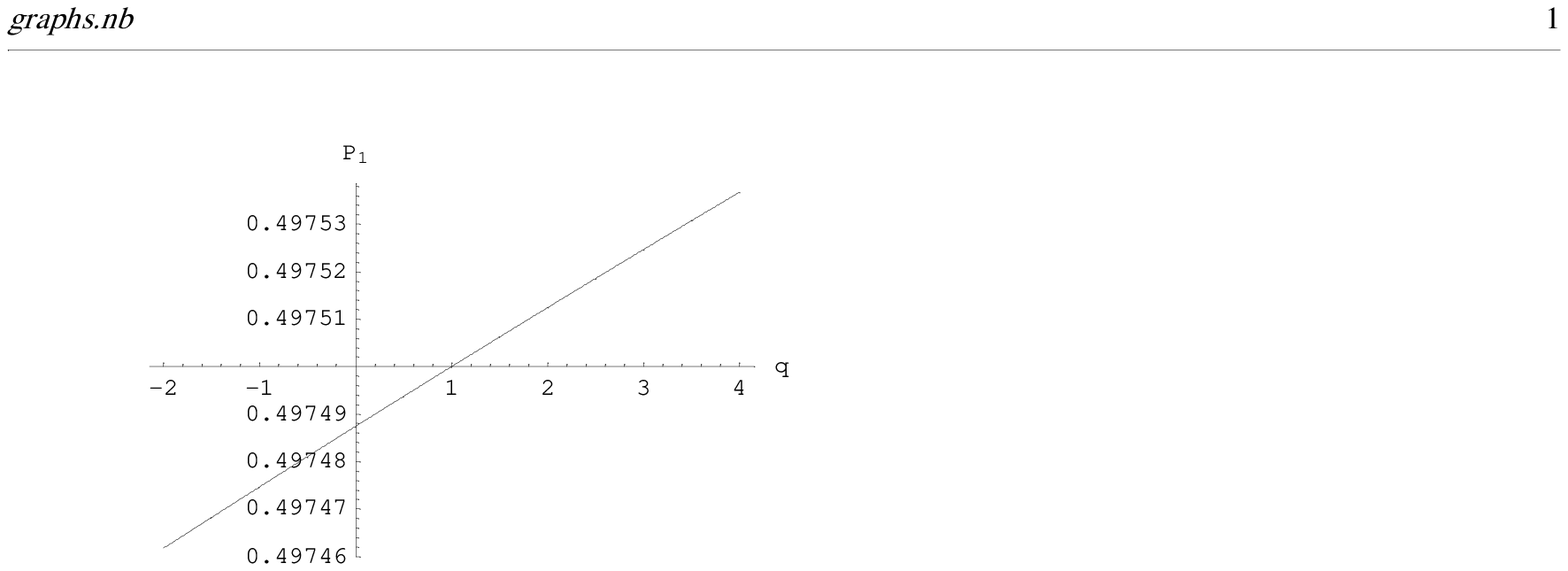}}
  \subfigure[at the value of $\beta \epsilon=4$]{\label{excitedprobability-b} \includegraphics*[viewport=125 565 300 708,clip]{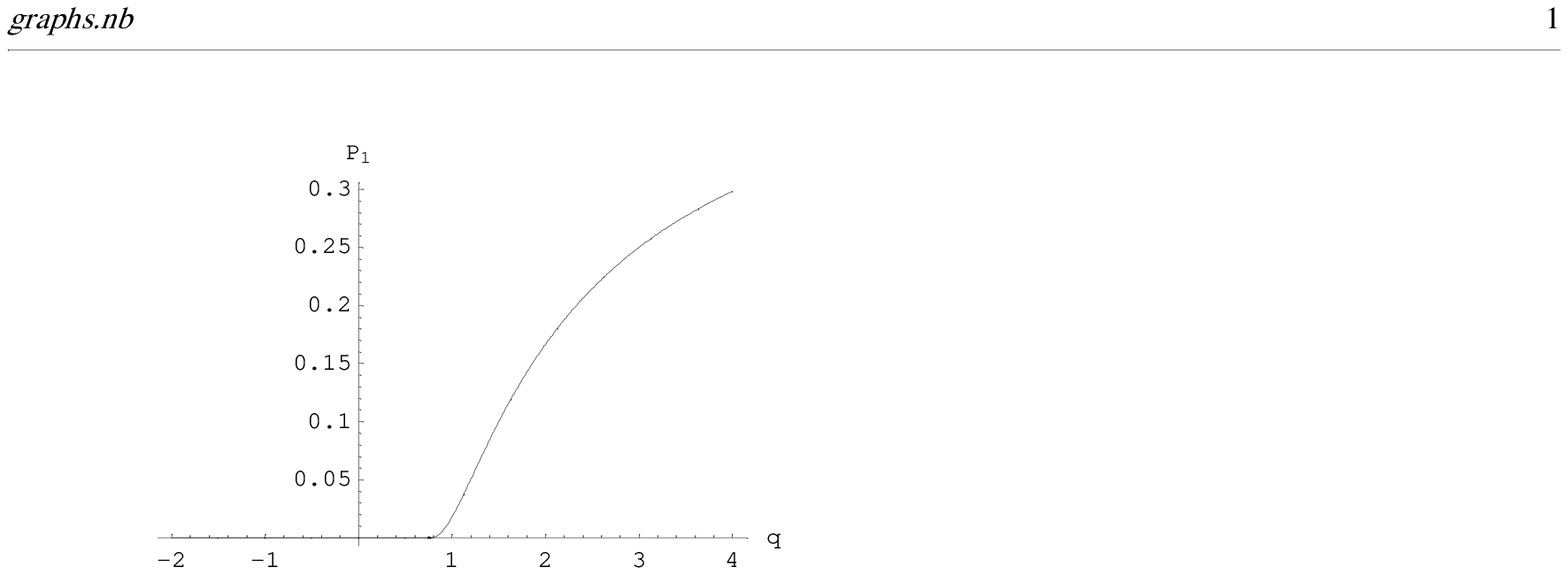}}
 \end{center}
\caption{The probability of a two-state system being in the
excited state as a function of $q$.}
\label{excitedprobability}
\end{figure}

It can be shown that the energy gap for a spin-$\frac{1}{2}$
system, in a magnetic field of the order of one Tesla, is of the
order of $10^{-23} \rm{J}$ for electrons and $10^{-27} \rm{J}$ for
nuclei. Thus, $\epsilon$ for this TSS has a value of
$10^{-25} \rm{J}$, yielding $\beta \epsilon=\frac{10^{-2}}{T}$.
Therefore, for a temperature range of 1 to 0.01 K, $\beta \epsilon$ ranges from 0.01 to 1. $P_1$ as a function of $q$
has been sketched in figure (\ref{excitedprobability}), for the
values of $\beta \epsilon$ equal to 0.01 and 4. In the first case,
for a unit change in $q$ the population of the
higher energy state undergoes a change of the order of $10^{-5}$,
while in the second case that change is of the order of $10^{-1}$. Thus, for a TSS the sensitivity to the value of $q$
increases by decreasing the temperature. For a typical value of
the energy separation between states, it seems impossible to
observe the effect of a change in $q$, unless considering very low
temperatures. Studying $S_q$ versus $q$ and $\beta \epsilon$ shows
that higher values of $q$ reduce the sensitivity of $S_q$ to
$\beta \epsilon$, and $q$ is again playing a role similar to
 temperature. This is a peculiar graph, since it
contains a number of peaks; its study is reserved for the future.

 For an harmonic oscillator, $\frac{h \nu}{k_B}$ ranges from 6215 for $\rm{H}_2$ to 133 for $\rm{K}_2$
\cite{mcquarrie}. Thus, $\Delta E=h \nu$ for the vibration of
a diatomic molecule is of the order of $10^{-20} \rm{J}$, resulting in $\beta
\epsilon = \frac{10^3}{T}$. Studying the
populations of the ground and first excited state versus $q$ at the values of $\beta \epsilon = 10, \;1,
\; \rm{and}\; 0.1$ shows that the sensitivity of the
population of the ground and first excited states to the value
of $q$ increases with decreasing the temperature.

\section{Rare event weight} \label{Rare}
It is claimed that, since the expectation value of an observable
is evaluated through $\langle A \rangle =\sum_i p_i^q A_i$, $q<1\;(q>1)$ privileges the rare (frequent) event \cite{abe:okamoto}. Since, in the present example $P_1$ is the
rare event and $P_0$ is the frequent event, figure (\ref{fig1}) shows the
opposite of the mentioned conclusion, that is because, $P_i$
itself is $q$-dependent. Thus, in order to make a valid judgement
regarding the effect of $q$ on rare or frequent events, we must
study
\begin{equation}\label{eventprobability}
p_i^q\propto \big[1-(1-q) \beta \epsilon_i \big]^{q/(1-q)}.
\end{equation}

To study (\ref{eventprobability}), the definition of rare (frequent) as the state with smaller (larger) probability lose its meaning. But, we can define the state with a larger (smaller) $\epsilon$ as the rare (frequent) event.
For large values of $q$, $p_i \propto \big[1-(1-q) \beta \epsilon_i
\big]^{-1}$, which is prefaring the frequent event. A numerical study of (\ref{eventprobability}) for small values of $q$ shows the privilege of rare events for negetive $q$'s (when they are allowed) and privilege of frequent events for positive $q$'s. The case of $q=0$ resembles the case of $T=0$ in Fermi-Dirac statistics, all states have the same weight, until the maximum value of $\beta \epsilon = 1$ is reached.
\section{The legitimate range of $q$} \label{legitimate}
In order to obtain physical properties of a system from its
partition function, the partition function must be a definite
function of the system's externally determined parameters.
Therefore, a partition function which is divergent does not
represent a physical system.
For an N-dimensional (D) harmonic oscillator with a single frequency,
$\nu$, the
partition function is
\begin{equation}\label{ndharmoscilpart1}
Z_q=\sum_{n=0}^\infty \frac{(N+n-1)!}{(N-1)!n!}\left[1+(q-1)\beta
h \nu \left(n+\frac{N}{2}\right)\right]^{\frac{1}{1-q}},
\end{equation}
where $n=\sum_i n_i$ is the number of excitons.
(Note that even in the absence of any interaction, the
multiplication of the single mode partition functions does not
yield to the overall partition function of the system.)  In the
limit of large $n$, the multiplicity (apart from the constant
$\frac{e^N}{N^{N-1}}$) behaves like $n^{N-1}$. Therefore, the
series converge for $q<1+\frac{1}{N }$. For the 1-D case it is
easy to use the integral test and consider the truncation of the
series to get $q<2$.

For a d-D particle in a box, by approximating the sum in the partition function as an
integral, we have
$Z_q \propto \int_0^{\infty}
\epsilon^{d/2-1} \big[1+(q-1)\beta \epsilon\big]^\frac{1}{1-q} d
\epsilon.$
The convergence condition for this integral is $1+\frac{2}{d}>q$. In the
case of 1-D, it is easy to show that the partition function is
convergent for $q<3$.

In the 2-D rigid rotor, $Z_q=\sum_{j=1}^\infty2\big[1-(1-q)\beta
\frac{\hbar^2}{2I}j^2\big]^\frac{1}{1-q}+1$. In the limit of large
$j$, the terms of the above series will behave like
$j^{\frac{2}{1-q}}$. Considering the range of $q$ where the series
is truncated, and using the integral test, we have $q < 3$ as the
acceptable range of $q$. In the 3-D rigid rotor,
$Z_q=\sum_{j=0}^\infty(2j+1)\big[1-(1-q)\beta
\frac{\hbar^2}{2I}j(j+1)\big]^\frac{1}{1-q}$. In the limit of
large $j$, the terms will behave like $j^{\frac{3-q}{1-q}}$;
Therefore, $q < 2$ yields a convergent partition function.

\section{Conclusion}
In non-extensive statistical mechanics there is a limitation
imposed on the values of $q$, due to the convergence of the
partition function series. By considering the results of section
(\ref{legitimate}), we can see that in an ideal gas, where $d
\rightarrow \infty$ or in a bath of harmonic oscillators where $N
\rightarrow \infty $, we have $q \leq 1$. Based on this observation, we
conjecture that in the thermodynamic limit, regardless of the
specific system under consideration, we must have $q \leq 1$. At the same time a large negative value of $q$ doesn't seem physical because it freezes the system in a few number of its lower energy levels. For
nano-systems the number of particles in the system is not so
large; thus, $q$ may be slightly larger than 1. This may be a starting
point for the study of nonextensivity in nano-systems.

Revisiting the common believe regarding the effect of $q$ on rare and frequent events show that the issue is more complicated than what is considered in the litrature. Large values of $q$ prefare the frequent event, but the situation is more complex for small values of $q$.

Physical properties of a system depend on the value of $q$ through the
population of states. The sensitivity of the population of states
to the value of $q$ decreases with increasing the temperature, for
some model systems. Therefore, it is possible for a system
believed to obey the BG statistics, to obey the Tsallis statistics
with a value of $q\ne 1$ but close to 1. This can be verifiable
only in infinitely low temperature experiments (which are not available).

We thank R. Kapral and E. Yazdian for useful discussions.

\bibliography{reference}
\end{document}